\documentclass[a4paper,10pt]{article}

\usepackage{amssymb,amsmath}
\usepackage{graphicx}
\usepackage{framed}
\usepackage{subfigure}

\renewcommand{\tfrac}[2]{#1/#2}

\begin{document}

\title{One-bit stochastic resonance storage device}

\author{S. A. Ib\'a\~nez${}^1$ \and P. I. Fierens${}^1$ \and G. A. Patterson${}^2$ \and R. P. J. Perazzo${}^1$ \and D. F. Grosz${}^{1,2,3}$}
\footnotetext[1] {Instituto Tecnol\'ogico de Buenos Aires (ITBA), Argentina.}
\footnotetext[2] {Departamento de F\'isica, FCEN, Universidad de Buenos Aires (UBA), Argentina.}
\footnotetext[3] {Consejo Nacional de Investigaciones Cient\'ificas y T\'ecnicas (CONICET), Argentina. \\
\noindent
We gratefully acknowledge financial support from ANPCyT under Project PICTO-ITBA 31176.}

\maketitle

\begin{abstract}

The increasing capacity of modern computers, driven by Moore's Law, is accompanied by smaller noise margins and higher error rates. In this paper we propose a memory device, consisting of a ring of two identical overdamped bistable forward-coupled oscillators, which may serve as a building block in a larger scale solution to this problem. We show that such a system is capable of storing one bit and its performance improves with the addition of noise. The proposed device can be regarded as asynchronous, in the sense that stored information can be retrieved at any time and, after a certain  synchronization time, the probability of erroneous retrieval does not depend on the interrogated oscillator. We characterize memory persistence time and show it to be maximized for the same noise range that both minimizes the probability of error and ensures synchronization.
We also present experimental results for a hard-wired version of the proposed memory, consisting of a loop of two Schmitt triggers. We show that this device is capable of storing one bit and does so more efficiently in the presence of noise.
\end{abstract}

\section{Introduction}
The increasing capacity of modern computers has been driven by Moore's Law, which postulates that the maximum number of transistors in an integrated circuit doubles every two years. However, as noted in \cite{PCMOS2008}, noise inmunity and power consumption do not follow Moore's law.
On the contrary, higher transistor density and power consumption require the use of smaller supply voltages. All these factors together lead to smaller noise margins and higher error rates in computation (\cite{Kish2004}). There have been several proposals to solve this problem, e.g., \cite{PCMOS2008,Palem2005,PCOMPUTING2003} take explicitly into account the fact that results may be correct only with some probability, and \cite{Kish2009} uses a set of orthogonal noise processes to represent logic values. Recently \cite{Murali2009}, it was shown how to implement basic logical operations (OR, AND, NOR, NAND) using nonlinear systems such that their performance improves in presence of noise, a signature of stochastic resonance.

Stochastic resonance (SR) is usually associated with a nonlinear system where the noise helps, an otherwise weak signal, to induce transitions between stable equilibrium states \cite{BENZI01,SR_Theory1,SR_Theory2}.
The phenomenon of stochastic resonance has been studied in a large number of applications, ranging from biological and neurological systems  \cite{wiesenfeld95, 608149, DBLP:journals/npl/RousseauC04}, information transmission sustained by noise \cite{Chapeau-Blondeau03,Chapeau-Blondeau04,PhysRevLett.81.5048,PhysRevLett.80.5212, Garcia-Ojalvo,German}, to information storage \cite{Memoria1,Memoria2,Memoria3}. In \cite{Memoria1,Memoria2}, a ring of identical oscillators was shown to be able to sustain a travelling wave with the aid of noise, long after the harmonic drive signal had been switched off. It is only natural to ask whether such a scheme can be used to store data, i.e., aperiodic signals, in noisy environments. In this work, we analyze the dynamic behavior of the shortest ring possible, comprising only two forward-coupled bistable oscillators. We show that such a system is capable of storing a single bit of information via stochastic resonance.

Memory performance is characterized in terms of the probability of an erroneous bit detection. In particular, we show that by the addition of a small amount of noise, the system outperforms the deterministic (noiseless) case. We also show that information can be retrieved from any of the two oscillators, obtaining the same probability of error after an elapsed `synchronization' time that decreases with increasing noise. Moreover, there is a noise range that yields a minimum probability of error with a nearly minimum synchronization time. By comparing system performance to that of the noiseless case, we define a memory persistence time and show it to exhibit a stochastic-resonance behavior.

Finally, we build a model of the proposed system with two Schmitt triggers (STs) in a loop configuration. STs provide a convenient `discrete' model of the bistable oscillators \cite{Carusela2003415}. By feeding each ST with Gaussian noise, we show that the system is capable of storing a single bit, and it does so more efficiently for an optimum amount of noise.

The paper is organized as follows: Section \ref{sec_thesystem} describes the system under analysis. In Section \ref{sec_simulationresults} we present simulation results and characterize the performance of the proposed memory device. Experimental results corresponding to the loop of Schmitt triggers are presented in Section \ref{sec_experimentalresults}. We close this work with the conclusions in Section \ref{sec_conclusions}.

\section{The system: a ring of two oscillators}
\label{sec_thesystem}

The system consists of a ring of bistable forward-coupled oscillators. The coupling is proportional to the amplitude of each oscillator, as it is described in \cite{Memoria1}. The system is described by the following set of  stochastic differential equations
\begin{eqnarray} \label{SistemaInicial}
dx_1 &=& \left[ -\frac{\partial U_{1}}{\partial x}\left(x_{1}\right)+\epsilon \frac{x_{N}}{x_0} \right]  dt + \sigma dW_{t,1},\nonumber\\
dx_n &=& \left[ -\frac{\partial U_{1}}{\partial x}\left(x_{n}\right) +\epsilon \frac{x_{n-1}}{x_0} \right]  dt + \sigma dW_{t,n}\:\:\:\:\:\:\:\:\:\:\:\:\:\:\:\:\: \mbox{for $1<n\leq N$},
\end{eqnarray}
where $N$ is the number of oscillators, $W_i$ represents standard Brownian motion, spatially uncorrelated ($\left\langle W_i \right\rangle =0$, and $\left\langle dW_i\,dW_j \right\rangle = \delta_{ij} dt$), $\sigma^2$ is the noise intensity, $\epsilon$ is the coupling strength between adjacent oscillators, and $U_1\left(x\right)$ is the one-dimensional potential defined by
\begin{equation}\label{Potencial1}
U_{1}\left(x\right) = U_0\left(\frac{x}{x_{0}} \right)^2\left[\left(\frac{x}{x_{0}} \right)^2-2 \right].
\end{equation}
We are interested in the shortest ring capable of storing information. We shall show that a loop of only two forward-coupled bistable oscillators not only presents an interesting dynamic behavior, but it can also work as a one-bit memory device. In the case of a ring comprising two oscillators, Eq. (\ref{SistemaInicial}) can be written as
\begin{equation}
d\vec{x} =  -\nabla U_{2}\left( \vec{x} \right) dt + \sigma d\overrightarrow{W},  \label{EcuacionMovimiento}
\end{equation}
where $d\overrightarrow{W}  = \left( dW_1 ,\,dW_2\right)$ is the noise vector and the bi-dimensional potential, $U_2\left( x_1,x_2\right)$, is expressed by
\begin{equation}
U_{2}\left(x_1,x_2\right) = U_0\left(\frac{x_1}{x_{0}} \right)^2\left[\left(\frac{x_1}{x_{0}} \right)^2-2 \right]+U_0\left(\frac{x_2}{x_{0}} \right)^2\left[\left(\frac{x_2}{x_{0}} \right)^2-2 \right]-\epsilon \frac{x_1 x_2}{x_0}.\label{Eq_Pot2}
\end{equation}
It is simple to show that Eq. (\ref{SistemaInicial}) cannot be written in the form of Eq. (\ref{EcuacionMovimiento}) for $N>2$. It is useful to understand Eq. (\ref{EcuacionMovimiento}) as the equation of motion of a particle in the bidimensional potential in Eq. (\ref{Eq_Pot2}).

\begin{figure}[t]
\centering
\includegraphics[scale=0.5]{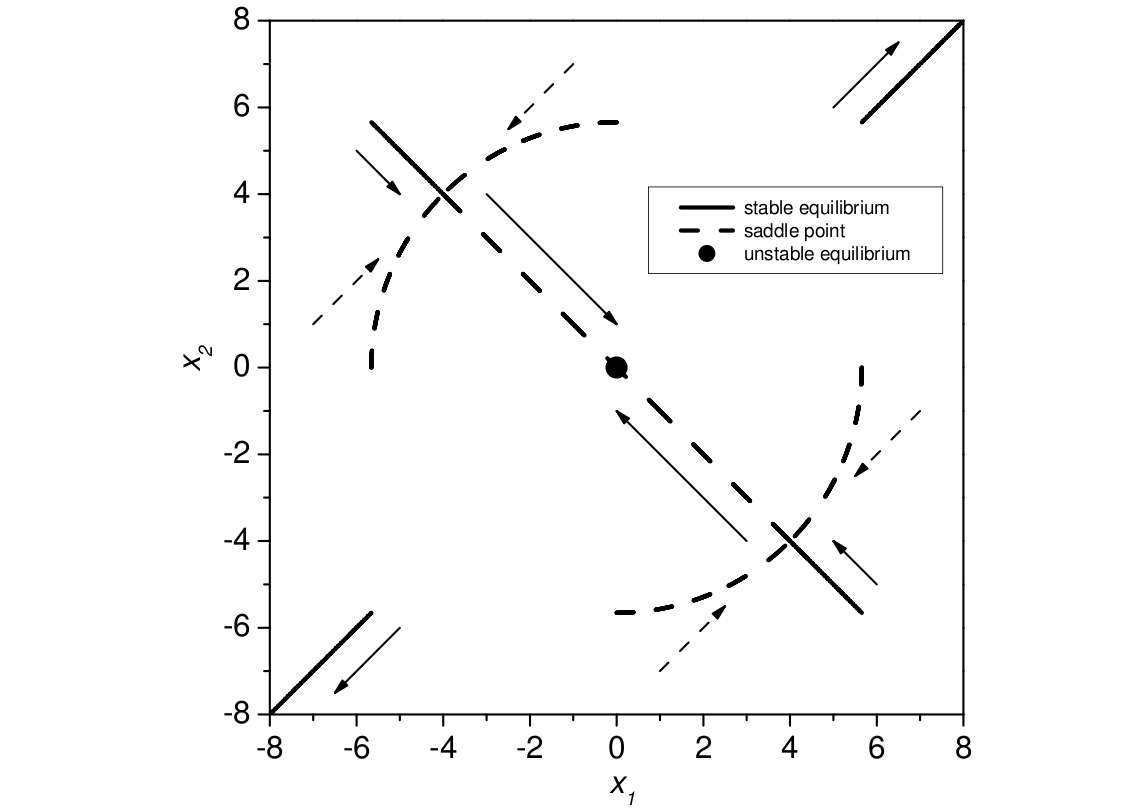}
\caption{Phase-space diagram for critical points of the system. Full lines correspond to stable equilibrium points, dashed lines correspond to saddle points and dotted lines correspond to unstable equilibria. The arrows indicate the direction  of increasing of $\epsilon$ .}
\label{Fig_EspFases}
\end{figure}

Depending on the coupling strength, the system, in the absence of noise, exhibits three different behaviors, and hence two bifurcations between them; the first at $\epsilon=\tfrac{2 U_0}{x_0}$ and the second at $\epsilon=\tfrac{4 U_0}{x_0}$. When $\epsilon<\tfrac{2U_0}{x_0}$, the potential in Eq. (\ref{Eq_Pot2}) presents one unstable equilibrium (i.e. a local maximum) at the origin $x_{1}^M=x_{2}^M=0$. It also exhibits four stable equilibrium points, one on each quadrant in the $x_1$-$x_2$ phase space (see Fig. \ref{Fig_EspFases}). While the equilibrium points located in the upper-right and lower-left quadrants are global minima, the equilibria located in the other two quadrants correspond to local minima. As the coupling strength $\epsilon$ increases, the wells corresponding to the global minima become deeper, and the wells of the local minima become shallower.

For $\epsilon = \tfrac{2 U_0}{x_0}$, the system undergoes a pitchfork bifurcation, where the local minima become saddle points, the equilibrium points located in the upper-right and lower-left quadrants remain global minima, and the origin remains a local maximum, as can be seen in Fig. \ref{Fig_EspFases}. There is another pitchfork bifurcation at $\epsilon = \tfrac{4 U_0}{x_0}$. In this case, the unstable equilibrium point $x_{1,M}=x_{2,M}=0$, develops into a saddle point. In this situation, the system is notably similar to the one-particle bistable potential in Kramers-Smoluchowski \cite{KRAMER}.

\begin{figure}[t]
\centering
\includegraphics[scale=0.4]{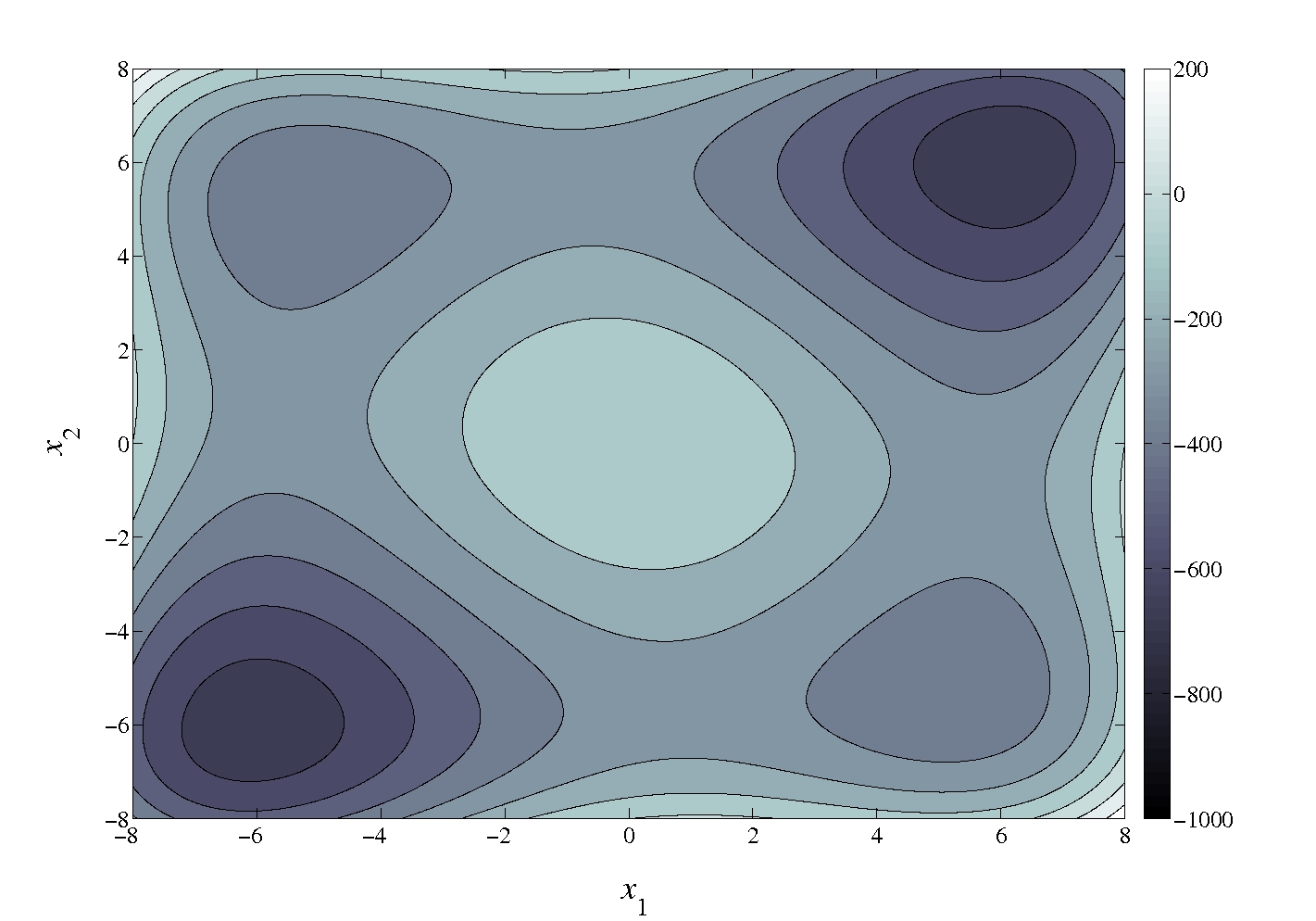}
\caption{The potential in Eq.(\ref{Eq_Pot2}) for $U_0=256$, $x_0=\sqrt{32}$ and $\epsilon=25$.}\label{Fig_Pot2}
\end{figure}

In this work, we focus solely on the region $0<\epsilon<\tfrac{2 U_0}{x_0}$ where the most interesting behavior is found. In Fig. \ref{Fig_Pot2}, we show the bidimensional potential in Eq. (\ref{Eq_Pot2}) for the particular case of $\epsilon \approx 0.56\,\tfrac{U_0}{x_0}$.

\subsection{Four-state model}

It is customary to approximate the behavior of a particle in a bistable potential by a two-state model (see, e.g., \cite{SR_Theory1,SR_Theory2}). Similarly, we can approximately describe the behavior of the system in Eq. \ref{EcuacionMovimiento} as that of a system with four discrete states, one for each equilibrium point of the potential in Eq. \ref{Eq_Pot2}. Let us enumerate the states as 0 through 3, starting from the equilibrium point in the  upper-right quadrant of Fig. \ref{Fig_EspFases} and moving in the counter-clockwise direction. If we denote by $n_i(t)$ the probability of finding the particle in the state $i$ at time $t$, then the behavior of the system can be described by the following equation
\begin{equation}
\dot{\vec{n}} = \mathbb{W} \cdot \vec{n},
\label{Eq_Maestra}
\end{equation}
where $\vec{n}= \left(n_0,n_1,n_2,n_3\right)^T$ and $\mathbb{W}$ is the transition matrix. We can estimate these transition rates as (see, e.g., \cite{KRAMER,Gardiner2004}) $W_{ij} \approx K_{ij} \exp\left\{\tfrac{-2\Delta U_{ij}}{\sigma^2}\right\}$, where $K_{ij}$ is a constant and $\Delta U_{ij}$ is the potential barrier that the particle has to overcome when moving from state $i$ to state $j$. Symmetries present in the potential in Eq. \ref{Eq_Pot2} (see Fig. \ref{Fig_Pot2}) allow us to reduce the problem to the calculation of four transition rates dependent on the following potential barriers
\begin{equation}
 \Delta U_{ij} = |j-i|U_0\left( 1+ (-1)^i\frac{\epsilon x_0}{2 |j-i| U_0} \right)^2,\;\;\text{for }ij = 01, 10, 13, 02.
  \label{Eq_Barreras}
\end{equation}
We shall come back to the four-state model in Section \ref{subsec_syncandpersistance}.

\section{Simulation results}
\label{sec_simulationresults}

In order to store one bit, we drive the first oscillator with a short pulse of amplitude $P_0$ and duration $T_P$. Since we are interested in a storage system that works solely in the presence of noise, we consider pulse intensities which are too low to force the system out of the deeper equilibrium states. In particular, for our simulations we set $P_0$ such that the work exerted by the driving force is smaller than the potential barrier $\Delta U_{01}$ in Eq. (\ref{Eq_Barreras}).

Memory retrieval from oscillator $i$, at interrogation time $t_0$, is performed by taking the average
\begin{equation}\label{Integrador}
\bar{x}_{i}(t_0)= \frac{1}{T_P}\int_{t_0}^{t_0+T_P}x_i(t) dt,
\end{equation}
and comparing $\bar{x}_{i}(t_0)$ to a fixed threshold, arbitrarily set to $0$. The system incurs in an error when $\bar{x}_{i}(t_0) < 0$.

We assess memory performance by observing the time evolution of the probability of error upon retrieval of the stored bit, $p_e$, estimated as the number of errors divided by the total number of realizations
\begin{equation}
p_e^i=\frac{\mbox{\# number of errors in oscillator $i$}}{\mbox{\# total number of realizations}},\;\text{for } i = 1,2.
\end{equation}

The results in this section correspond to a system with $U_0=256$ and $x_0=\sqrt{32}$. The pulse duration was set to $T_P=5$ and pulse amplitude to $P_{0}=0.13 \Delta U_{01} / x_{eq}$, where $x_{eq}=x_0 \sqrt{1+\tfrac{\epsilon x_0}{4 U_0}}$ is the distance from the global minima to the coordinate axes.  For each noise level, we performed $10000$ numerical simulations of Eq. (\ref{EcuacionMovimiento}), with initial conditions chosen randomly in the region $\left[-9.0:+9.0 \right] \times \left[-9.0:+9.0 \right]$, using the Euler-Maruyama method \cite{EulerMaruyama} with the integration step $\Delta t=6.1035\times10^{-4}$.

\begin{figure}[t]
\centering
\subfigure[$p_e$ vs. time for $\epsilon=75$]{
\label{Fig_BERvsTIME_a}
\includegraphics[scale=0.30]{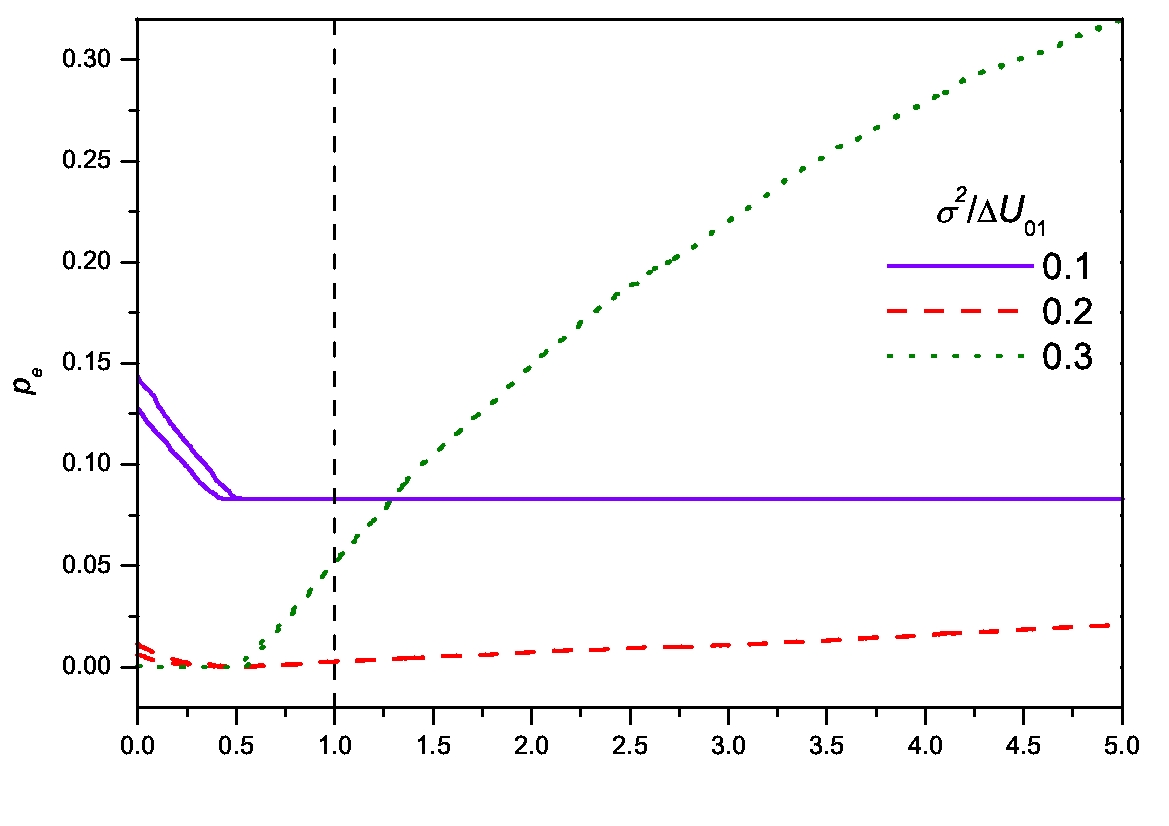}
}
\subfigure[$p_e$ vs. time for $\epsilon=25$]{
\label{Fig_BERvsTIME_b}
\includegraphics[scale=0.30]{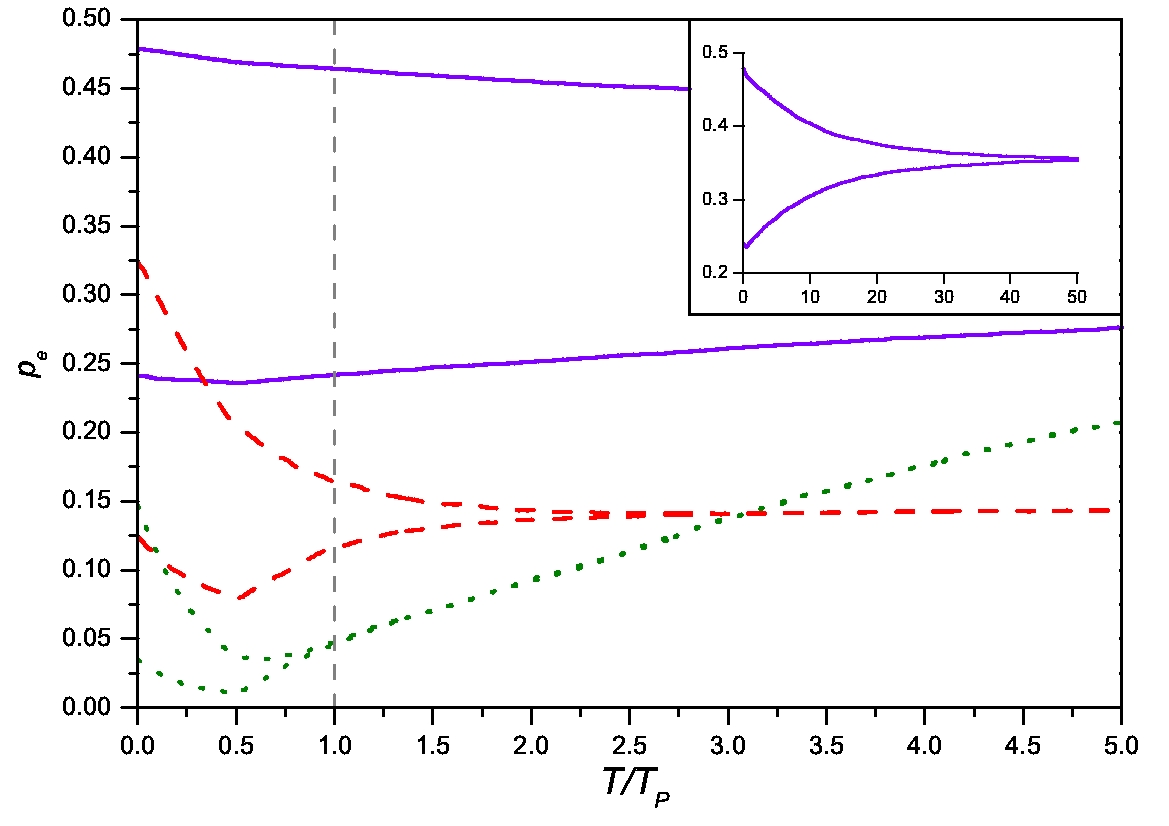}
}
\caption{The evolution of the probability of error as a function of noise intensity and coupling strength. In both figures, $p_e^1$ is always smaller than $p_e^2$.
}
\label{Fig_BERvsTIME}
\end{figure}

In Fig. \ref{Fig_BERvsTIME} we show the evolution of the probability of error as a function of noise intensity and coupling strength. The dotted line indicates the instant at which the drive is turned off. Observe that the probability of error for the first oscillator is always smaller than that for the second. In the case of strong coupling (Fig. \ref{Fig_BERvsTIME_a}), the probability of error improves with time in both oscillators while the driving pulse is on. For small noise intensity, the degradation of the error probability is almost negligible after turning off the pulse, since noise is not strong enough to overcome the potential barrier. On the other hand, performance degrades at a faster pace for higher noise intensities. While the behavior of both oscillators is very similar for a strong coupling, for weaker coupling and a small noise intensity, the first oscillator outperforms the second (see Fig. \ref{Fig_BERvsTIME_b}). Indeed, the first oscillator rapidly follows the driving pulse, which is stronger than the weakly-coupled output of the second oscillator. However, the second oscillator is driven by the sub-threshold output of the first oscillator and, thus is not able to immediately follow the external drive. When the driving pulse is turned off $p_e^1$ slowly increases because of noise and the influence of the second oscillator. On increasing the noise intensity, the performance of both oscillators becomes similar independently of the coupling strength.

\begin{figure}[t]
\centering
\subfigure[$p_e^1$ vs. noise for $\epsilon=75$]{
\label{Fig_BER_Sig_a}
\includegraphics[scale=0.24]{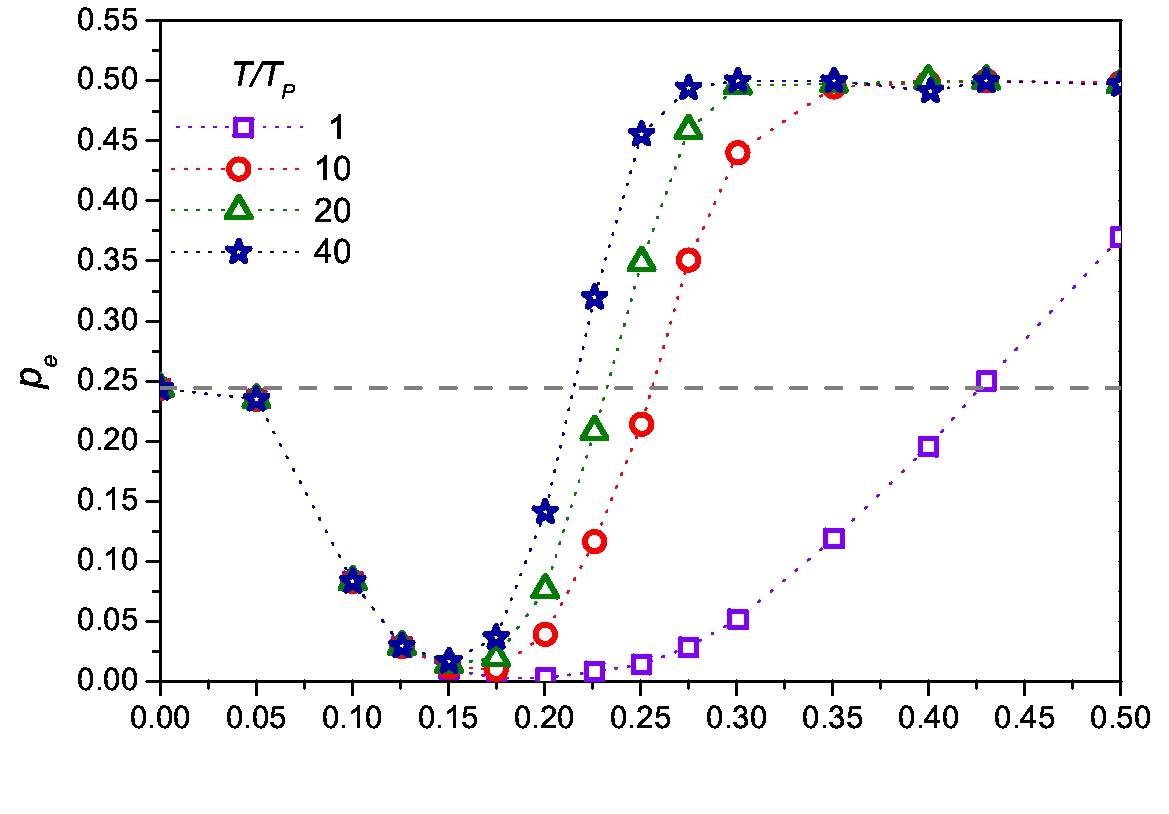}
}
\subfigure[$p_e^2$ vs. noise for $\epsilon=75$]{
\label{Fig_BER_Sig_b}
\includegraphics[scale=0.24]{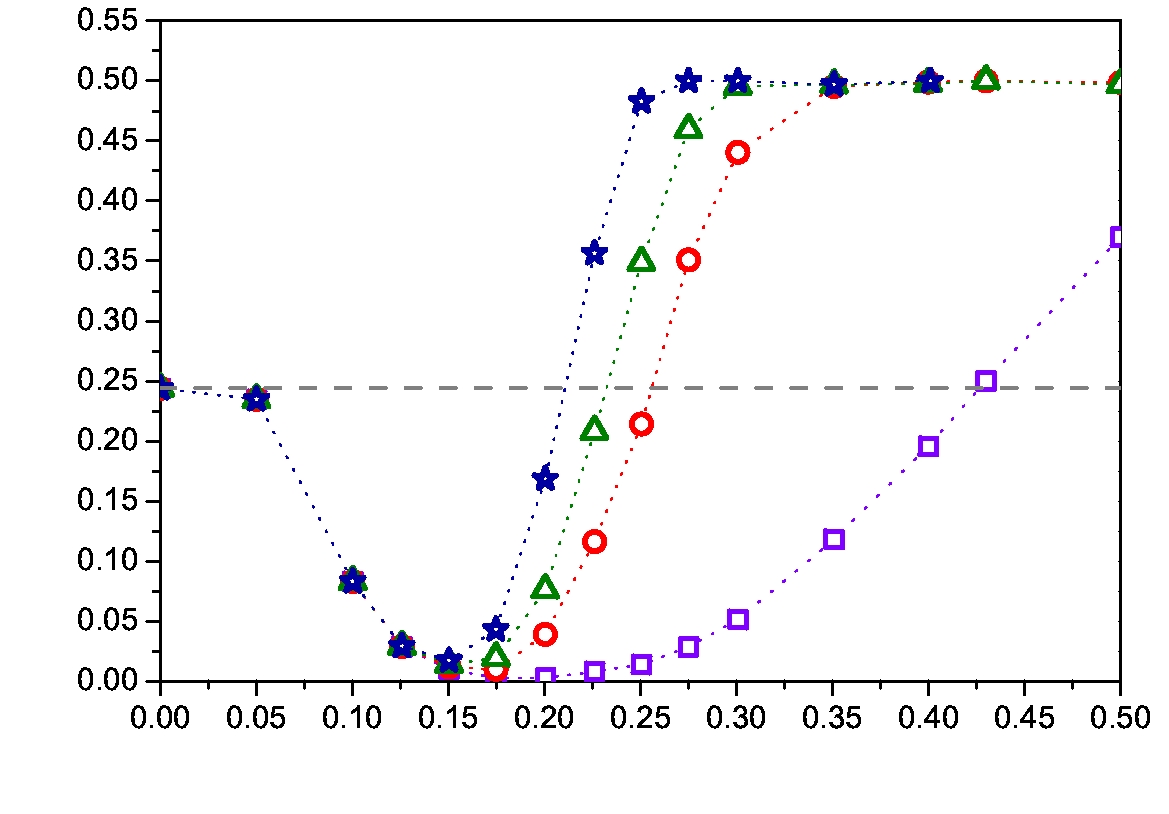}
}
\subfigure[$p_e^1$ vs. noise for $\epsilon=25$]{
\label{Fig_BER_Sig_c}
\includegraphics[scale=0.24]{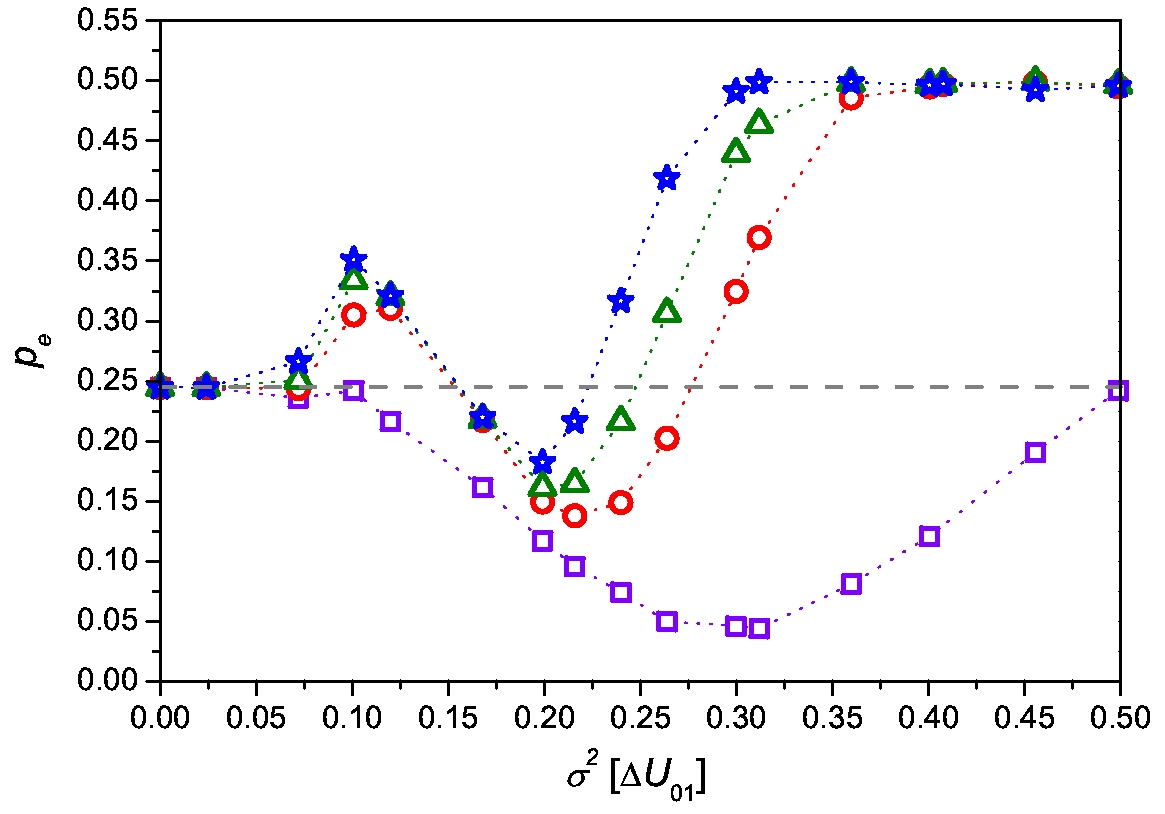}
}
\subfigure[$p_e^2$ vs. noise for $\epsilon=25$]{
\label{Fig_BER_Sig_d}
\includegraphics[scale=0.24]{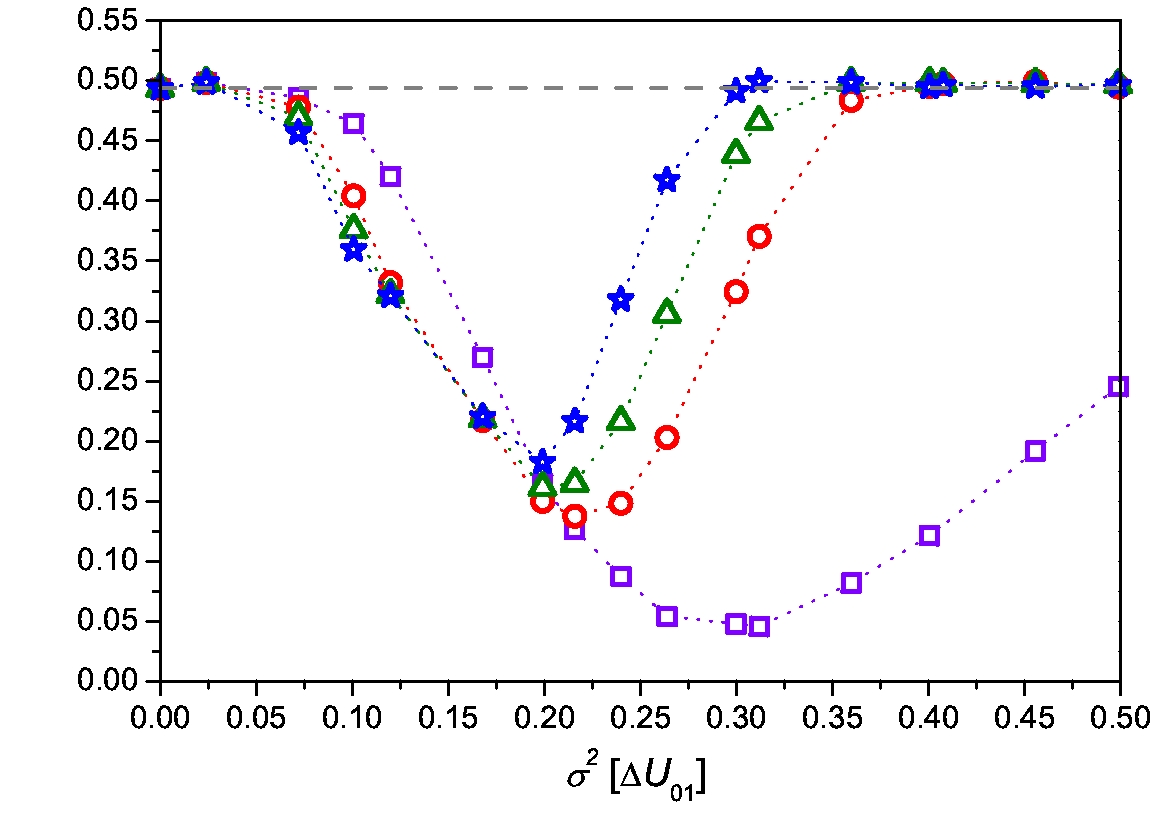}
}
\caption{Probability of error as function of noise intensity for different observation times ($T=T_P$, $10 T_P$, $20T_P$, $40T_P$) and coupling strengths ($\epsilon = 25$ and $75$).}
\label{Fig_Ber_Sig}
\end{figure}

Figure \ref{Fig_Ber_Sig} shows the estimated probability of error for each oscilllator as a function of noise intensity and for increasing times ($T_1=T_P$, $T_2=10\,T_P$, $T_3=20\,T_P$, and $T_4=40\,T_P$). It is interesting to compare these results to the deterministic noiseless system. In this case it is simple to estimate error probabilities. Consider the phase space in Fig. \ref{Fig_EspFases}. Since the driving pulse is only applied to the first oscillator, a particle in the upper-right and lower-right quadrants remains there after the pulse is turned off. Moreover, the pulse is strong enough to move a particle from the upper-left to the upper-right quadrant, but it is too weak to force the particle out of the lower-left quadrant (where there is a deep minimum of the potential, as shown in Fig. \ref{Fig_Pot2}). Therefore, the probability of error for the first oscillator is only $1/4$. However, the probability of error for the second oscillator depends on the coupling strength. If the coupling is strong, then the output of the first oscillator is capable of driving the second and $p_e^2 = 1/4$. On the other hand, if the coupling strength is weak, the state of the second oscillator does not change as a consequence of changes in the output of the first and, hence, $p_e^2 = 1/2$. A small amount of noise is enough to improve memory performance as it allows a particle located in the lower-right quadrant to escape to adjacent wells. Fig. \ref{Fig_Ber_Sig} shows the positive influence of noise. An antiresonant behavior is readily observed, i.e., the probability of error is optimized for a given noise intensity, a signature of stochastic resonance.

Now we turn to the dynamic behavior of the coupled oscillators. Results shown in Fig. \ref{Fig_Ber_Sig} indicate that the coupling strength plays a significant role in the dynamics of the memory device: for weak coupling strengths,  the minimum $p_e$ increases with time (i.e, memory performance degrades), whereas it remains approximately constant for a strong coupling strength. This is a desirable feature for implementation of a practical device, as it points to a system whose performance, not only benefits from added noise, but it is also robust in a noisy environment. As expected, in both cases the noise range that yields the minimum $p_e$ decreases with time.

\subsection{Synchronization and memory persistence}
\label{subsec_syncandpersistance}

Keeping in mind a practical realization of the coupled double-well system as a 1-bit storage device, not only we require information retrieval to be `asynchronous' in the sense that it is possible to interrogate the system at any time, but also that we obtain a unique value for the memory state when we interrogate any of the two oscillators. We say that the oscillators are `synchronized' if $p_e^1$ and $p_e^2$ differ in a small quantity, arbitrarily fixed to $10^{-3}$. Results in Fig. \ref{Fig_BERvsTIME_b} show the time evolution of the probability of error for three different noise intensities, starting at the time when the external pulse is switched on, until the two oscillators are synchronized. We define the elapsed time between these two events as the  synchronization time $T_s$. By comparing the performance of the two oscillators, we find that, until synchronization is reached, the second oscillator exhibits a worse error rate for all noise levels. This is due to the fact that the external pulse acts exclusively on the first oscillator.

\begin{figure}[t]
\centering
\includegraphics[scale=0.40]{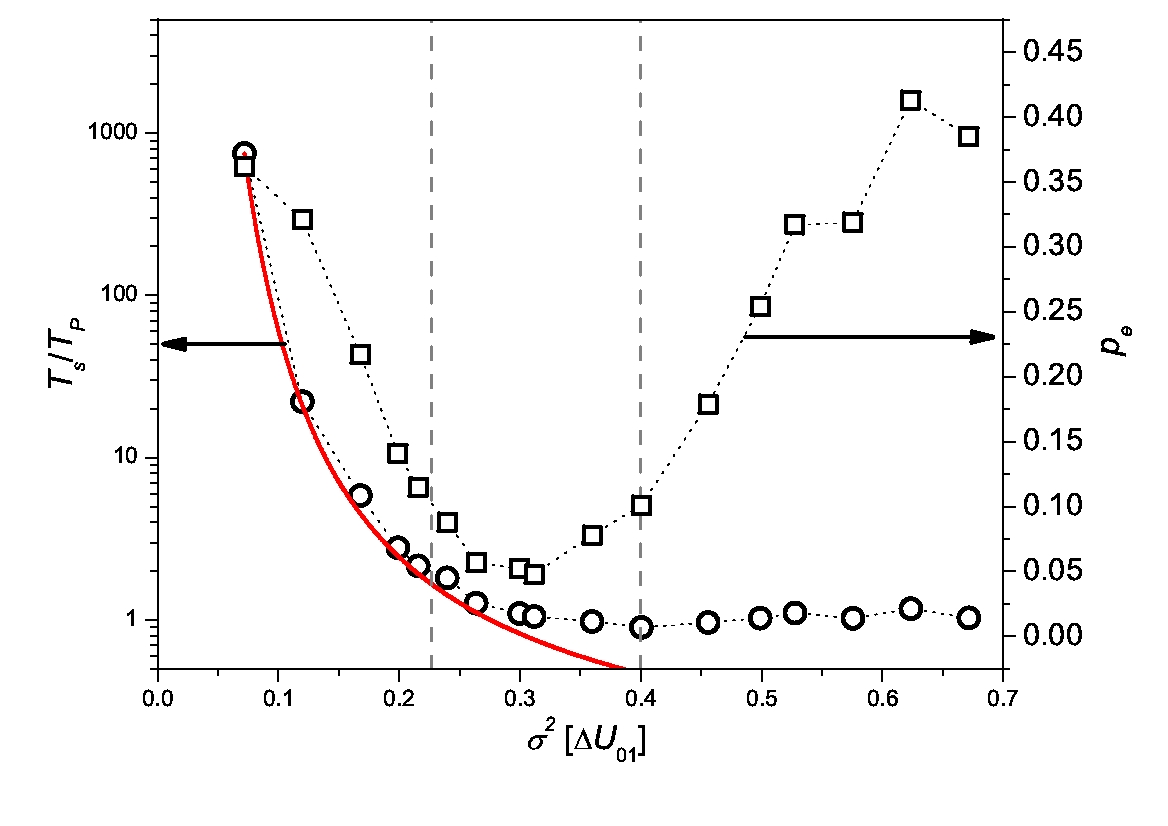}
\caption{Synchronization time ($T_s$) and probability of error at  $T_s$ as a function of noise intensity.  Solid line corresponds to a theoretical approximation to $T_s$.}
\label{Fig_ACOP}
\end{figure}

Fig. \ref{Fig_ACOP} shows the synchronization features (i.e. synchronization time and probability of error at synchronization) as a function of the noise intensity. Interestingly, increasing the noise intensity makes $T_s$ decrease.

The behavior of the synchronization time when it is greater than $T_P$ can be qualitatively described as follows: suppose that memory retrieval is carried out by instantaneously observing the state of the oscillator, i.e., without computing the average in Eq. (\ref{Integrador}). Then, the probability of error can be written in terms of the state probabilities in Eq. (\ref{Eq_Maestra}) as $p_e^1(t) = n_1(t)+n_2(t)$ and $p_e^2(t) = n_2(t)+n_3(t)$. Solving Eq. (\ref{Eq_Maestra}) for a given initial condition, it can be shown that
\[
 p_e^1(t)-p_e^2(t) \propto \mathrm{e}^{-2 (W_{10} + W_{13}) t}.
\]
Therefore, the synchronization time must be
\begin{equation}
T_s = \frac{\tau}{2 (W_{10} + W_{13})},
\label{Eq_TsTeo}
\end{equation}
where $\tau$ is a suitably chosen constant. Although we have derived Eq. (\ref{Eq_TsTeo}) assuming instantaneous observations of the oscillators, it is reasonable to expect that the equation is still approximately valid when the average in Eq. (\ref{Integrador}) is computed for $T_s > T_P$. Indeed, we can observe that Eq. \ref{Eq_TsTeo} fits well the numerical results in Fig. \ref{Fig_ACOP}, where $W_{10}$ and $W_{13}$ were computed using Eq. (\ref{Eq_Barreras}) and $\tau$ was calculated by fitting only the simulated value corresponding to the smallest noise intensity.

As it can be observed in Fig. \ref{Fig_ACOP}, there is a range of noise intensities,  enclosed by the dotted lines, for which both $p_e$ is small and  $T_s$ is close to its minimum, i.e, the device performs well and any of the two oscillators can be interrogated, at any time, in order to retrieve the memory state.

\begin{figure}[t]
\centering
\includegraphics[scale=0.40]{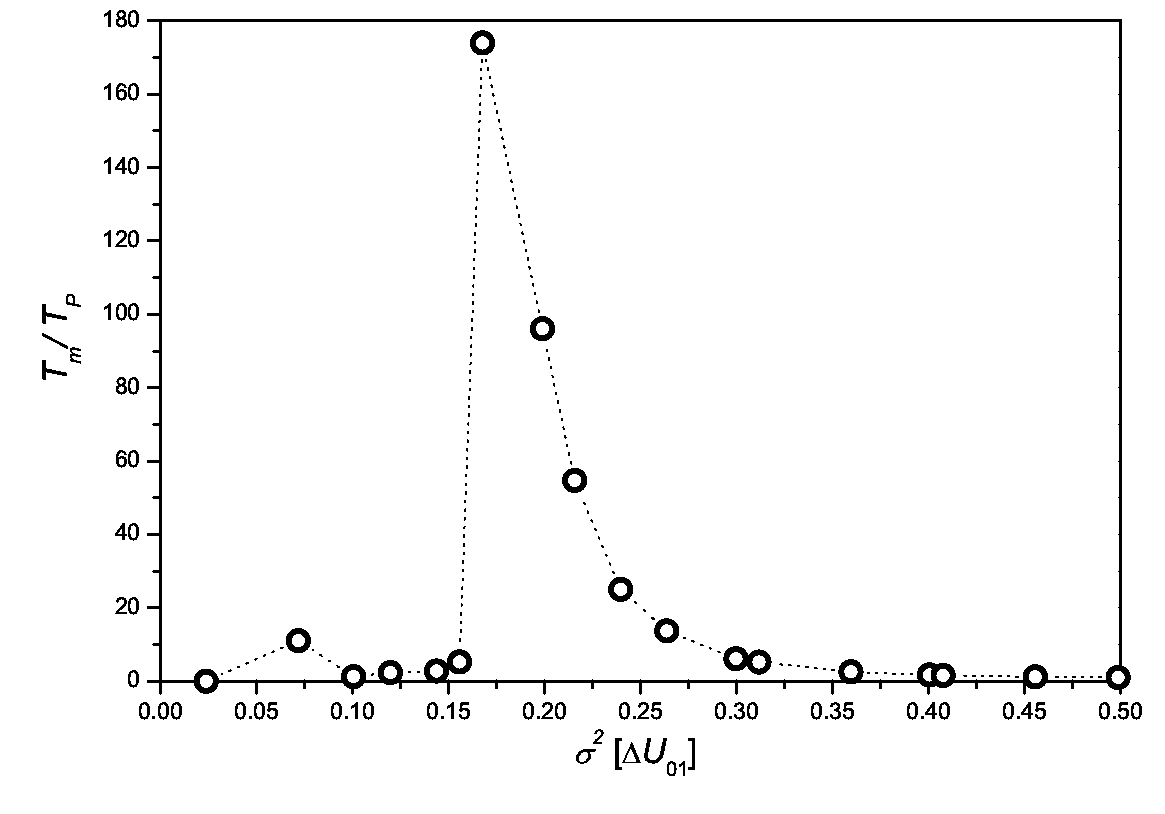}
\caption{Persistence time as a function of noise intensity.}
\label{Fig_LifeTime}
\end{figure}

Finally, we propose the following criterion as a way to characterize the persistence time of the memory, ($T_m$): we take ($T_m$) as the time elapsed until the first oscillator reaches a probability of error rate equal to that corresponding to the noiseless case, i.e, from that point on, noise no longer helps improve device performance. This definition allows us to specify a `refresh' time scale, in a similar fashion as in common DRAM devices (see, e.g., \cite{DRAMThesis}).  Note that, from Fig. \ref{Fig_LifeTime} where $T_m$ is shown as a function of noise intensity and for a weak coupling strength, $T_m$ presents a stochastic resonance characteristic. The explanation of this behavior lies in the fact that, for an optimal noise level, the system moves easily to the upper-right equilibrium state, but hardly ever leaves it. Note that for this optimal noise level, \textit{the memory device not only maximizes the persistence time, but also both oscillators are synchronized and the probability of error is minimal.}

\section{Experimental results}
\label{sec_experimentalresults}

In this section we present experimental results corresponding to a ring of two Schmitt triggers. We use STs as `discrete' models of the bistable potentials in previous sections. 
A single pulse, representing the `1' state that is to be stored, is fed into the first ST with a supra-threshold amplitude of +5 V and duration $T_P = 1$ ms. The ST thresholds are set to +3 V and -1 V, respectively. The output of the first ST, set to sub-threshold values of +2 V and 0 V respectively, is used to drive the second ST. Then, the output of the second ST is fed back into the first ST and we allow for a delay of $t = 98 T_P$ before refreshing the memory by again applying the driving pulse. In the experimental setup, noise is generated by low-pass filtering of a pseudo-random bit sequence (PRBS) generator working at a rate of 250 kHz \cite{Horowitz}. Since the filter cutoff 3-dB bandwidth is chosen greater than 10 KHz, the noise spectral density is flat over the studied range.

During the time when the system is not driven, we interrogate the second ST at intervals of $T_P$. As discussed in previous sections, the detector averages the received amplitude over a fraction of $T_P$ and compares it to a fixed threshold, in order to make a decision between a `1' and a `0' states. Finally, we repeat this procedure 1000 times, alternating the initial state of the second ST, and compute the probability of error by counting the number of times a `0' state was detected.

\begin{figure}[t]
\centering
\includegraphics[scale=0.40]{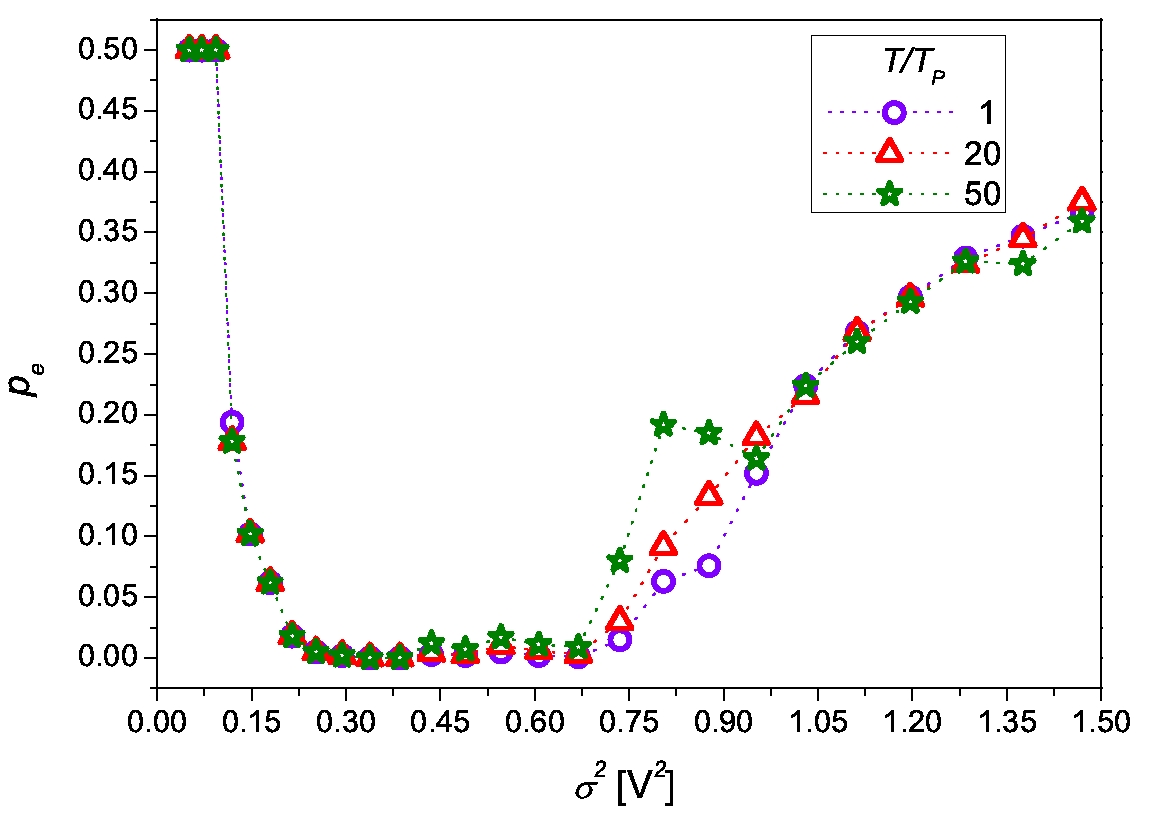}
\caption{Experimental results for a loop of two STs: Memory performance as a function of the noise intensity and time.}
\label{Fig:PerformanceSTs}
\end{figure}

Fig. \ref{Fig:PerformanceSTs} shows the estimated probability of error as a function of noise intensity and elapsed time. As it can be observed, $p_e$ behaves in a similar manner as in Fig. \ref{Fig_Ber_Sig}(d). For instance, $p_e = 1/2$ for low noise intensities because the sub-threshold output of the first ST is not strong enough to drive the second ST. It can also be observed that there is a range of noise that minimizes the probability of error, a signature of stochastic resonance which is also present in the results in Fig. \ref{Fig_Ber_Sig}. In this range we computed no errors, i.e., estimated $p_e < 10^{-3}$. Also, around the optimum noise intensity, device performance does not degrade with time.
In summary, experimental results obtained with discrete bistable potentials built around Schmitt triggers reproduce well the main characteritics of the proposed stochastic-resonance 1-bit memory device, paving the way for a practical implementation.

\section{Conclusions}
\label{sec_conclusions}

 In this work, we studied the dynamical behavior of a system comprised of two bistable oscillators in a loop configuration. In particular, we showed that such a system is capable of storing a single bit of information in a noisy environment. Furthermore, by calculating the probability of error upon retrieval of the stored bit, we showed that performance in the presence of noise exceeds that of the deterministic noiseless system, a signature of stochastic resonance.

The proposed device can be regarded as `asynchronous', in the sense that stored information can be retrieved at any time. Moreover, after a certain  `synchronization' time, the probability of erroneous retrieval does not depend on the interrogated oscillator. Interestingly, we found that there is a range of noise intensities which both minimizes the probability of error and the synchronization time.

 System performance can also be characterized by the memory persistence time, which we defined as the time elapsed until the probability of error equals that of the noiseless case. We found that this persistence time is maximized for the same noise range that minimizes the probability of error and ensures synchronization.

 Then, we built a model of the proposed device by means of a loop of two  Schmitt triggers, acting as `discrete' analogues of the bistable oscillators. We were able to show experimentally that this system is capable of storing a single bit and does so more efficiently in the presence of noise.

In summary, we believe that the proposed device may serve as a building block of future computing systems which, due to the increasing scale of integration, will have to deal with smaller noise margins.

\bibliographystyle{unsrt}
\bibliography{two-osc-mem}

\end{document}